\documentclass{article}
\newcommand{\pp}{$p+p$}
\newcommand{\AuAu}{$\rm Au+Au$}
\newcommand{\pT}{$p_{T}$}
\newcommand{\sqrtsnn}{$\sqrt{s_{_{\mathrm{NN}}}}=200$ GeV}
\newcommand{\sigmastar}{\mbox{$\Sigma ^{*} (1385)$}}
\newcommand{\sigmastarbar}{\mbox{$\overline{\Sigma ^{*}} (1385)$}}

\newcommand{\gs}{\mbox{$\gamma_{s}$}}
\newcommand{\mB}{\mbox{$\mu_{B}$}}
\usepackage{lajolla2006}
\usepackage{graphicx}
\frompage{000} \topage{000}                                              
\def\rightmark{Statistical Models and STAR's Strange Data}
\def\leftmark{S. Salur et al.}

\title{Statistical Models and STAR's Strange Data }
\authors{
{Sevil Salur (for the STAR Collaboration)
}\\[2.812mm]
{\normalsize Yale University, 272 Whitney Ave., New Haven, CT 06520 USA\\[0.2ex]
}}

\abstract{The yields of strange hadrons have been measured as a
function of centrality in \AuAu\ and in \pp\ collisions at
$\sqrt{s_{NN}}=200$ GeV in STAR. The system size and energy
dependence are studied and compared for \pp\ and \AuAu\
collisions. Thermal models are fitted to the ratios of various
strange particles to investigate the particle production and to
determine the strangeness enhancement. The temperatures ($T$) and
the strangeness enhancement factors (\gs) of the systems
determined from the fits are presented.}

\PACS{specifications see, e.g.\ {\tt http://www.aip.org/pacs/}}

\begin{document}

\maketitle
\setcounter{page}{1}

\section{Introduction}\label{intro}

It was pointed out by Fermi and later by Hagedorn that particle
production  in heavy ion collisions can be described with the
considerations of phase space \cite{ref:hagedorn2}. The question
that is being discussed is whether phase space arguments can also
describe the centrality dependence of strangeness production. The
differences in the production of bulk matter (\pT\ $<$ 1 GeV) and
non-bulk matter  (\pT\ $>$ 1 GeV) are investigated with
statistical models which are also used to estimate the equilibrium
properties and the trends of particle yields and ratios. The
quality of the statistical fits and the implications of the
resulting fit parameters such as $T$, \gs\ and \mB\ are discussed.

\section{Statistical  Models}\label{techno}

Assuming QGP formation in heavy ion collisions at RHIC energies,
it is expected that the thermal nature of the partonic medium
could be preserved during hadronization \cite{ref:PBraun}. The
particle yields measured in the final state then resemble a
population in thermal equilibrium. Thermal models are used to
predict the equilibrium properties of a macroscopic system from
the measured yields of the constituent particles. It is commonly
agreed that light (u and d) quarks are more likely to reach
equilibrium in the medium than the strange quarks due to the
relatively larger strange quark mass ($m_{s} \sim T_{c}$). The
strangeness saturation factor, $\gamma_{s}$\index{$\gamma_{s}$},
is introduced to account for the amount of strangeness chemical
equilibration \cite{ref:rafelskigammas}. One of the parameters of
the statistical fits, which is the more generic term of the
strangeness saturation factor (\gs), is the phase space occupancy
($\gamma_{i}$). This term is used to regulate the sum of particle
and anti-particle pairs produced in the medium.

With the given ratios of particles, it is possible to deduce the
temperature T\index{T}, the volume V\index{V}, the baryonic
chemical potential $\rm \mu_{B}$\index{$\rm \mu_{B}$}, strangeness
chemical potential $\rm \mu_{S}$\index{$\rm \mu_{S}$} and charge
chemical potential $\rm \mu_{Q}$\index{$\rm \mu_{Q}$}
\cite{ref:PBraun,ref:torrierithesis,ref:rafelskiqm}. Chemical
potentials describe the particle anti-particle difference and
require that the variables are conserved only on average in the
whole system (Grand Canonical (GC) ensemble). The GC ensemble can
be used to describe large systems such as central \AuAu\
collisions, while the Canonical ensemble is used for small system
such as \pp\ collisions when all quantum numbers must be conserved
exactly.

The predicting power of statistical models for T, $\mu_{B}$,
$\mu_{S}$, and $\gamma_{S}$ require the utilization of measured
particle ratios. With the ratios, all degeneracy factors of the
fireball cancel, leaving just the relative fugacities. Some
examples are given in Equation~\ref{eq:proton} and
Equation~\ref{eq:lambdafuga},

\begin{equation}\label{eq:proton}
    \frac{\bar{p}}{p} = \lambda^{-4}_{u}
    \lambda^{-2}_{d}=\exp(-(\frac{4\mu_{u}+2\mu_{d}}{T})),
\end{equation}

\begin{equation}\label{eq:lambdafuga}
    \frac{\Lambda}{p} = \lambda_{s} \lambda^{-1}_{u}=\exp(\frac{\mu_{s}-\mu_{u}}{T})
\end{equation}
where the fugacity of a hadron is defined by the product of its
valance quark fugacities (e.g.,
$\lambda_{\pi}=\lambda_{u}\lambda_{d}$,
$\lambda_{n}=\lambda_{u}\lambda^{2}_{d}$ and
$\lambda_{\Lambda}=\lambda_{u}\lambda_{d}\lambda_{s}$).

\section{Statistical Model Comparisons}\label{details}

Several statistical model computational codes are available. A
comparison of the parameters of these codes is presented in
Table~\ref{tab1}. In order to make any comparisons of the the
fits, the requirements of the models are fixed to be same.
\begin{table}[b] \vspace*{-12pt} \caption[]{Thermal Model
Comparisons \cite{ref:share,ref:ther}; their ensembles and the
free parameters that are used in the fits.}\label{tab1}
\vspace{0.12in} \vspace*{-14pt}
\begin{center}
\begin{tabular}{llllll}
\hline\\[-10pt]
Models Used  & Ensemble & Parameters \\
\hline\\[-10pt]\vspace{0.05in}
4 Parameter Fit& Grand Canonical &T, $\mu_{q}$, $\mu_{s}$, and
$\gamma_{s}$ \\\vspace{0.05in}
SHARE V1.2 &Grand Canonical & T, $\lambda_{q}$, $\lambda_{s}$,
$\gamma_{q}$, $\gamma_{s}$, $\mu_{I3}$, N, $\lambda_{c}$ and
$\gamma_{c}$ \\
THERMUS V2 & Canonical & T, B, S, Q, $\gamma_{s}$ and R  \\
 & Grand Canonical & T, $\mu_{B}$, $\mu_{s}$, $\mu_{Q}$, $\mu_{c}$, $\gamma_{s}$, $\gamma_{c}$ and R  \\
\hline
\end{tabular}
\end{center}
\end{table}
For example, $\gamma_{q}$ (light-quark phase space occupancy) is
fixed to 1 in SHARE \cite{ref:share} and only the Grand Canonical
ensemble is used in THERMUS \cite{ref:ther}. Also, since the
default contribution of the feed-down from weak decays on particle
ratios is handled differently in each code, either all the
feed-down is removed or is included in all models in the same
amounts.

\section{Statistical Model Fits}\label{maths}

\begin{figure}[here!]
\centering $\begin{array}{cc} 
\hspace{-0.3in}
  \includegraphics[height=6.1cm,clip]{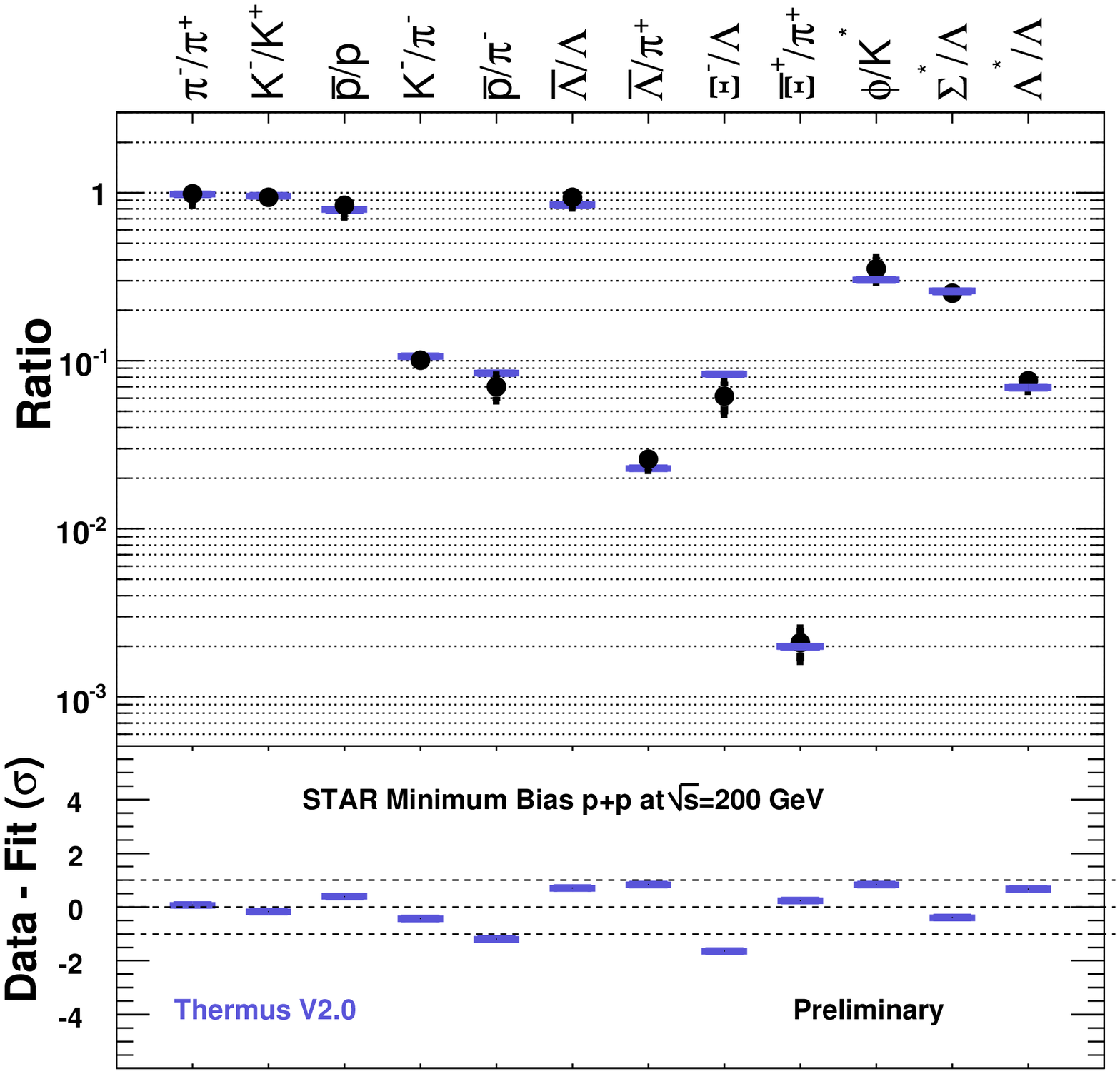} &
  \hspace{-0.1in}
\includegraphics[height=6.1cm,clip]{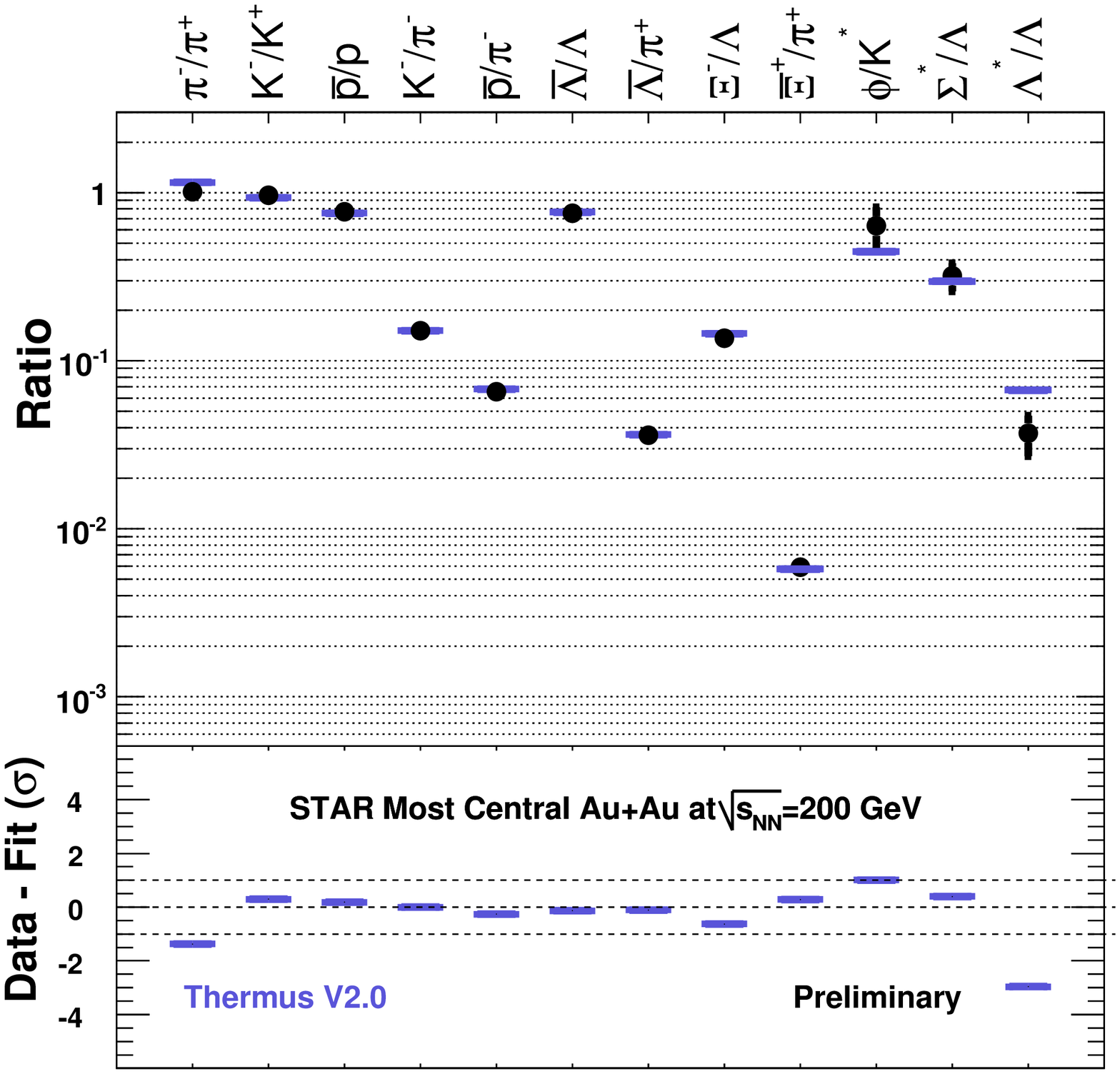} \\
 \mbox{\bf   (a) }&\mbox{\bf   (b)}
\end{array}$
\caption { {\bf   (a)}  Particle ratios in $\sqrt{s_{NN}}=200 $
GeV \pp\ collisions in comparison to canonical thermal model
predictions (short lines)  from Thermus V2.0 \cite{ref:ther}. The
fit parameters are $T_{ch}=171\pm9$ MeV, $\gamma_{s}=0.53\pm0.04$
and $r=3.49\pm0.97$ fm. B and Q are fixed to 2 while S is fixed to
0. {\bf   (b)} Particle ratios in the $\sqrt{s_{NN}}=200 $ GeV
most central \AuAu\ collisions in comparison to a grand canonical
thermal model prediction. The fit parameters are $T_{ch}=168\pm6$
MeV, $\gamma_{s}=0.92\pm0.06$, $\mu_{B}=(4.52\pm0.98)\times
10^{-2}$ GeV, $\mu_{S}=(2.23\pm0.74)\times 10^{-2}$ GeV,
$\mu_{Q}=(-2.05\pm0.77)\times 10^{-2}$ GeV, and $r=15\pm10$ fm.
See text for details.}
\label{fig:thermal}      
\end{figure}

Predictions of a thermal model are compared to the measured
particle ratios from \pp\ collisions at \sqrtsnn\ in
Figure~\ref{fig:thermal}-a.  The resulting fit parameters are
$T_{ch}=171\pm9$ MeV, $\gamma_{s}=0.53\pm0.04$
\index{$\gamma_{s}$} and the radius of the source, $r$, is
$3.49\pm0.97$ fm. The quantum numbers baryon (B) and charge (Q)
are fixed to 2 while net strangeness (S) is fixed to 0. The short
lines represent the thermal model fits from a canonical
calculation performed by Thermus V2.0. The closed circles
represent the particle ratios measured in p+p collisions. The
model accurately describes both the stable particle and resonance
ratios in p+p collisions. The 1 $\sigma$ error on the ratios and
the differences between the model predictions and the data are
included in the plot. The differences do not exceed 2 $\sigma$.

Figure~\ref{fig:thermal}-b presents the particle ratios and
Thermus V2.0 predictions for the most central \AuAu\ collisions at
\sqrtsnn. The measured ratios are used to predict the free
parameters by using a grand canonical approach. The fit parameters
are $T_{ch}=168\pm6$ MeV, $\gamma_{s}=0.92\pm0.06$,
$\mu_{B}=(4.52\pm0.98)\times 10^{-2}$ GeV,
$\mu_{S}=(2.23\pm0.74)\times 10^{-2}$ GeV,
$\mu_{Q}=(-2.05\pm0.77)\times 10^{-2}$ GeV, and $r=15\pm10$ fm.
The model describes all but two particle ratios within 1 $\sigma$
standard deviation. Except for the $\Lambda(1520)/\Lambda$ ratio,
the other measured resonance ratios ($\phi/K^{*}$,
$\sigmastar/\Lambda$) are within a 1 $\sigma$ error of the model
fits. Due to the very short lifetime ($\tau <
\tau_{fireball}\sim\;10$ fm) of most resonances, a large fraction
of their decays occur before the thermal freeze-out. Elastic
interactions of resonance decay products with particles in the
medium alters the momenta of these particles. This results in a
loss of the signal reconstructed for resonances (due to
re-scattering). However, secondary interactions (regeneration) can
increase the resonance yield. The effects of regeneration  and
re-scattering are not included in this thermal model which might
explain why the experimental values differ from the statistical
model calculations for the $\Lambda(1520)/\Lambda$ ratio. It is
expected that there should be no re-scattering and regeneration
effects on the $\phi$ meson due to its long life time. Since the
model predicts the ratios for \sigmastar\ and $K^{*}$ correctly,
re-scattering appears to be balanced with the regeneration
effects.

\begin{figure}[here!]
\centering $\begin{array}{cc} 
  \includegraphics[height=6cm,clip]{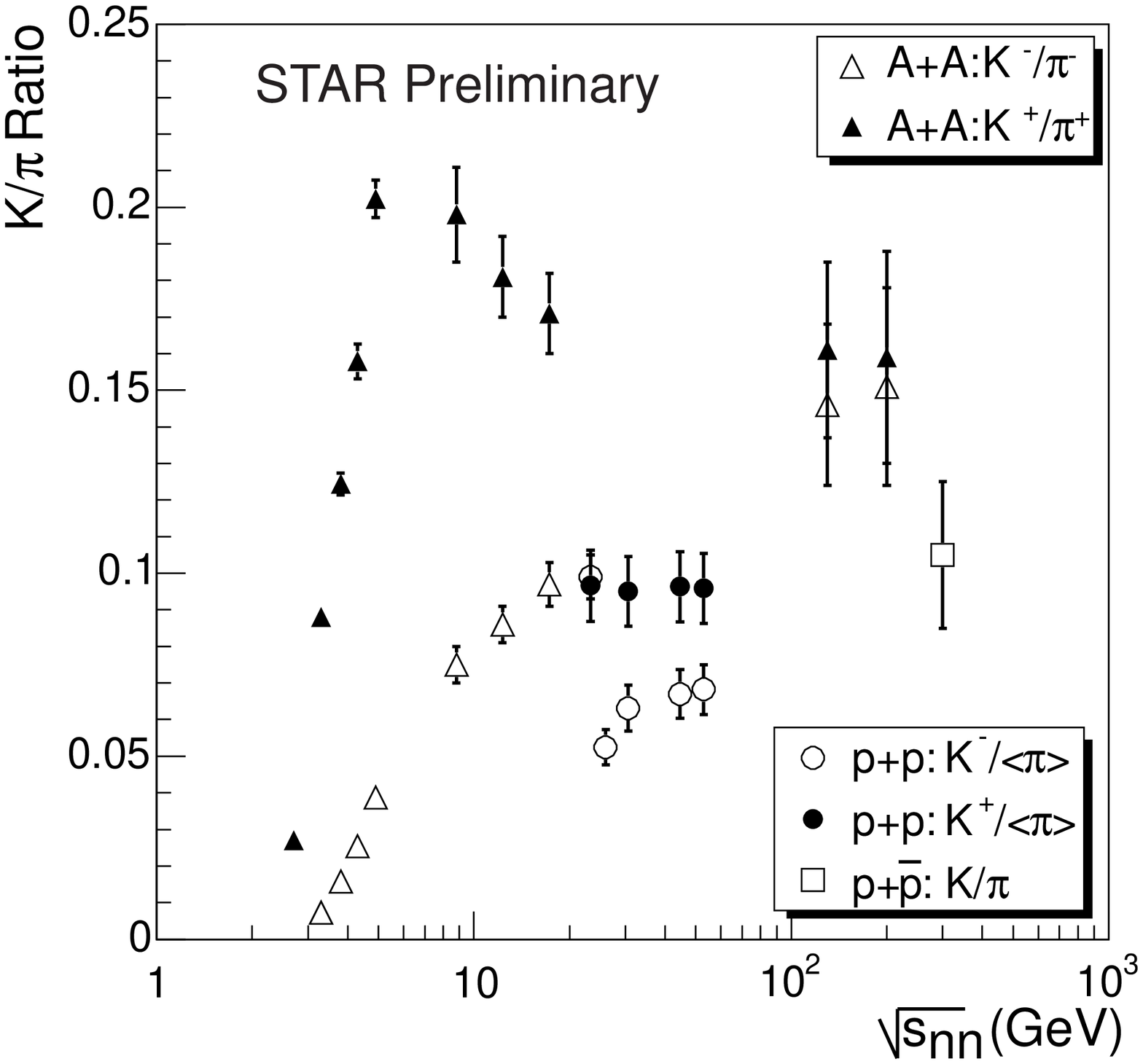} &
  \hspace{-0.1in}
\includegraphics[height=5.8cm,clip]{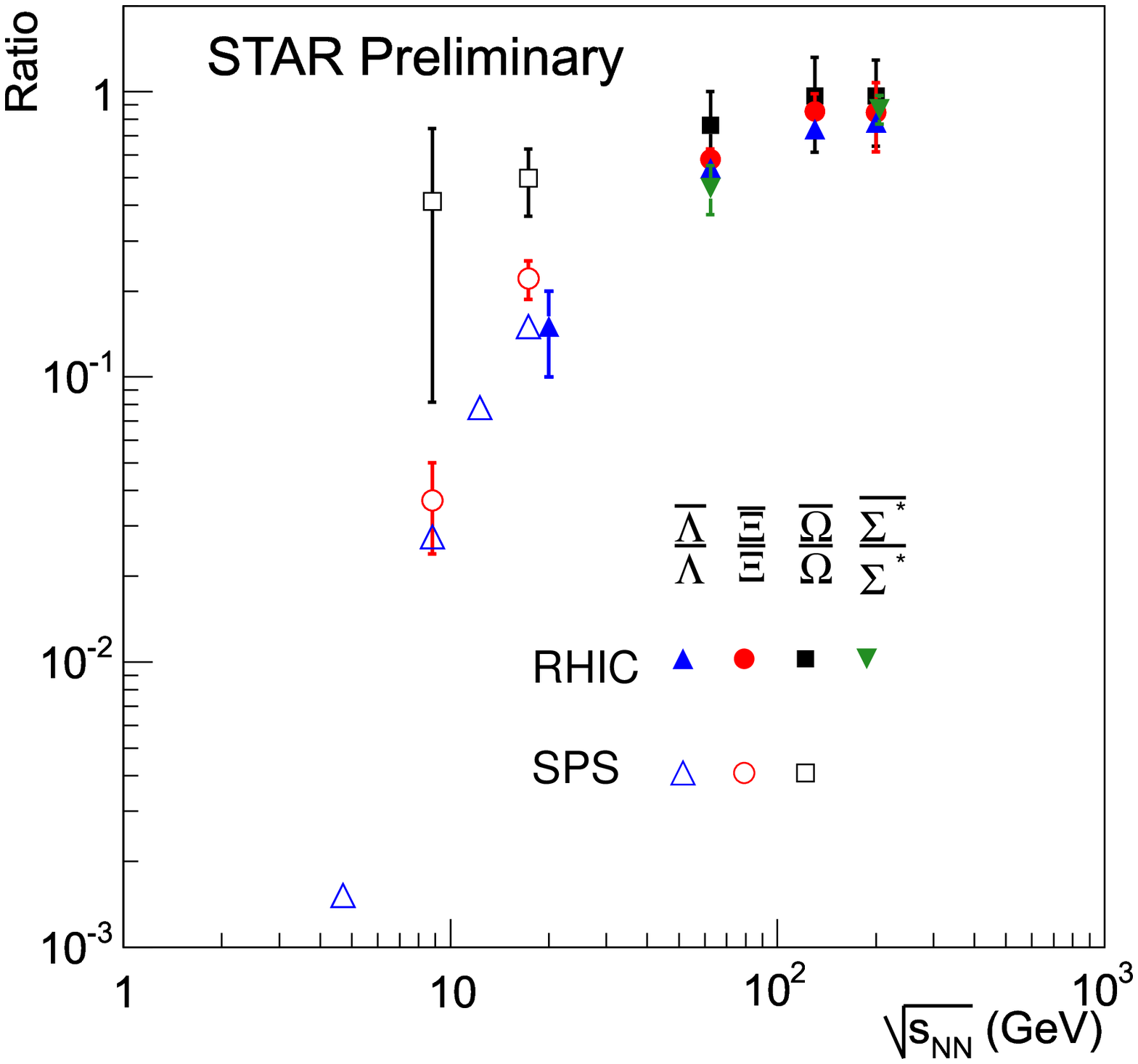} \\
 \mbox{\bf   (a) }&\mbox{\bf   (b)}
\end{array}$
\caption { {\bf   (a)} Mid-rapidity $K/\pi$ ratios versus $
\sqrt{s_{NN}}$
   and their dependence on heavy-ion and \pp\ collisions.  {\bf   (b)} The energy dependence of  the
   ratio of anti-strange to strange baryons from heavy-ion collisions (Pb+Pb for SPS \& \AuAu\ for RHIC).}
\label{fig:kpiratio}      
\end{figure}

In \AuAu\ collisions the free fit parameter corresponding to
strangeness saturation,  $\gamma_{s}$ \index{$\gamma_{s}$}, is
higher than in \pp\ collisions, while $T$ \index{T} is the same
within the errors. In Figure~\ref{fig:kpiratio}-a, the energy and
system size dependence of the measured $K/\pi$ ratios are
presented \cite{kpiratio}. An enhancement is observed in the
K/$\pi$ ratios in central Au+Au collisions relative to \pp\
collisions implying an increase in strangeness production
resulting in larger values of $\gamma_{s}$. The higher
$\gamma_{s}$ in \AuAu\ relative to \pp\ collisions is in agreement
with the $K/\pi$ ratio observations. Due to this increase of
strangeness production from \pp\ to \AuAu\ collisions, particle
ratios are selected such that $\mu_{s}$ cancels out (e.g.,
$\sigmastar/\Lambda$ ratio). The small value of the chemical
potential, $\mu_{B}$, can be explained by the proximity of the
measured anti-particle to particle ratios (e.g., $\sigmastarbar /
\sigmastar$ \cite{ref:salurthesis})  to unity and it reflects the
near zero net baryon number at mid-rapidity of \AuAu\ collisions.

 For comparisons with heavy-ion
systems, Figure~\ref{fig:kpiratio}-a shows the energy and system
dependence of the $K/\pi$ ratios at mid-rapidity. Triangles
correspond to data from heavy ion collisions and circles to \pp\
collisions at the given
energies~\cite{ref:ppallkaons,ref:ppkaons}. An enhancement in the
$K/\pi$ ratios of about 50\% is observed at RHIC energies in
central Au+Au collisions relative to \pp\ collisions extrapolated
to similar energies. A similar magnitude of enhancement in $
K^{-}/\pi$ has already been observed at the lower energies of the
AGS and SPS \cite{ref:agskaon,ref:spskaon}. 

The energy dependence of the $\Lambda$ and $\overline{\Lambda}$
yields at mid-rapidity from \AuAu\ collisions at RHIC and $\rm
Pb+Pb$ collisions at SPS as a function of $ \sqrt{s_{NN}}$ is
presented in Figure~\ref{fig:kpiratio}-b \cite{ref:salurQM04}.
From SPS to RHIC energies, strange baryon production is
approximately constant at mid-rapidity, whereas the
$\overline{\Lambda}$ rises steeply, reaching $80 \%$ of the
$\Lambda$ yield at RHIC top energies. The other hyperons, $\Xi$
and $\Omega$, follow similar trends. Since most of the strange
baryons produced also include light up and down quarks, strange
baryon production at low energies is dominated by valence quark
transport from the colliding system, but at RHIC it is dominated
by pair production. This also implies that at RHIC energies, most
of the incoming baryons continue moving with large rapidity and do
not cause any increase in the baryon number in the mid-rapidity
regions.

\section{Statistical Coalescence?}\label{others}

\begin{figure}[here!]
\centering $\begin{array}{cc} 
\hspace{-0.1in}
  \includegraphics[height=6.1cm,clip]{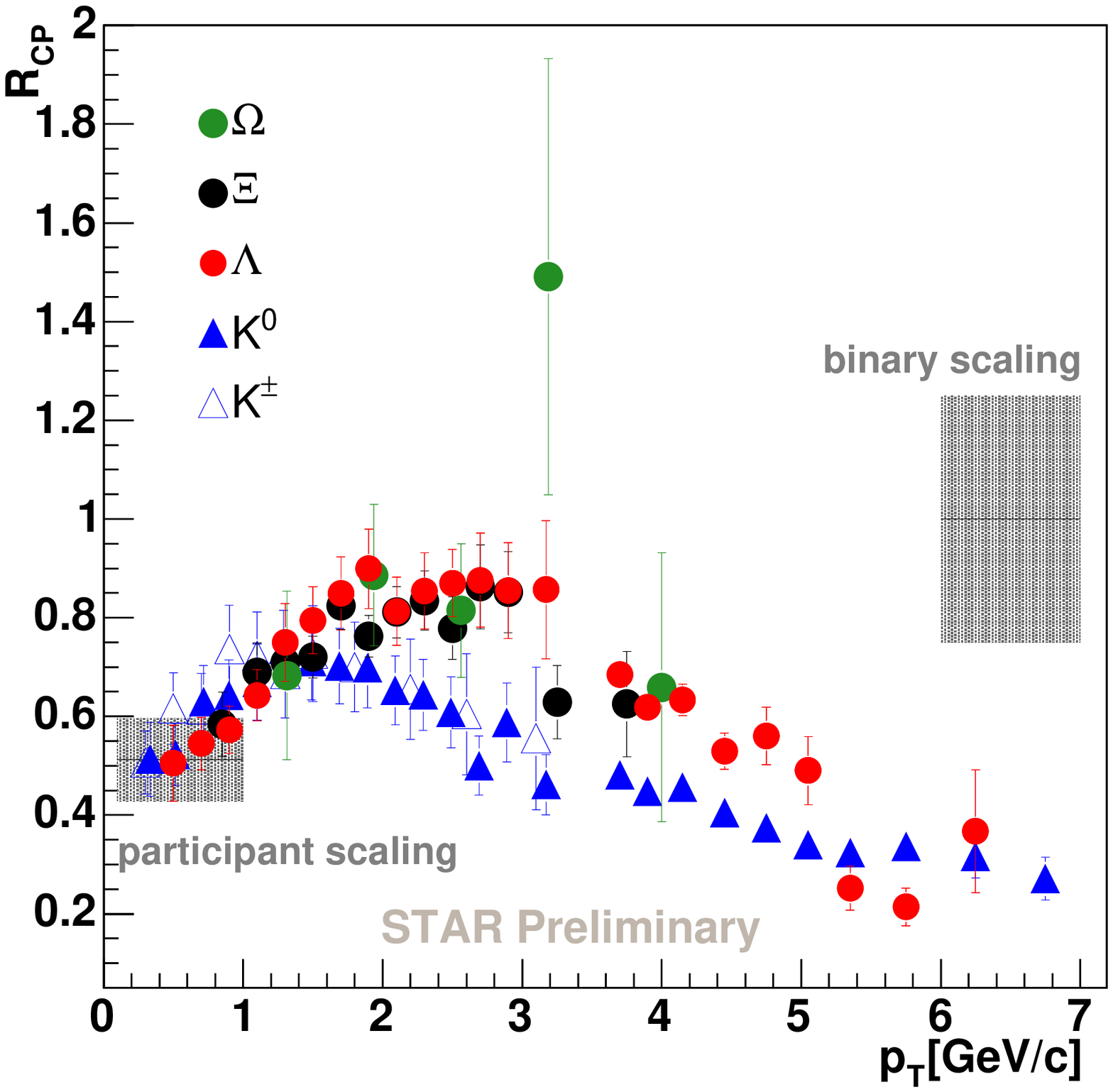} &
  \hspace{-0.1in}
\includegraphics[height=6.1cm,clip]{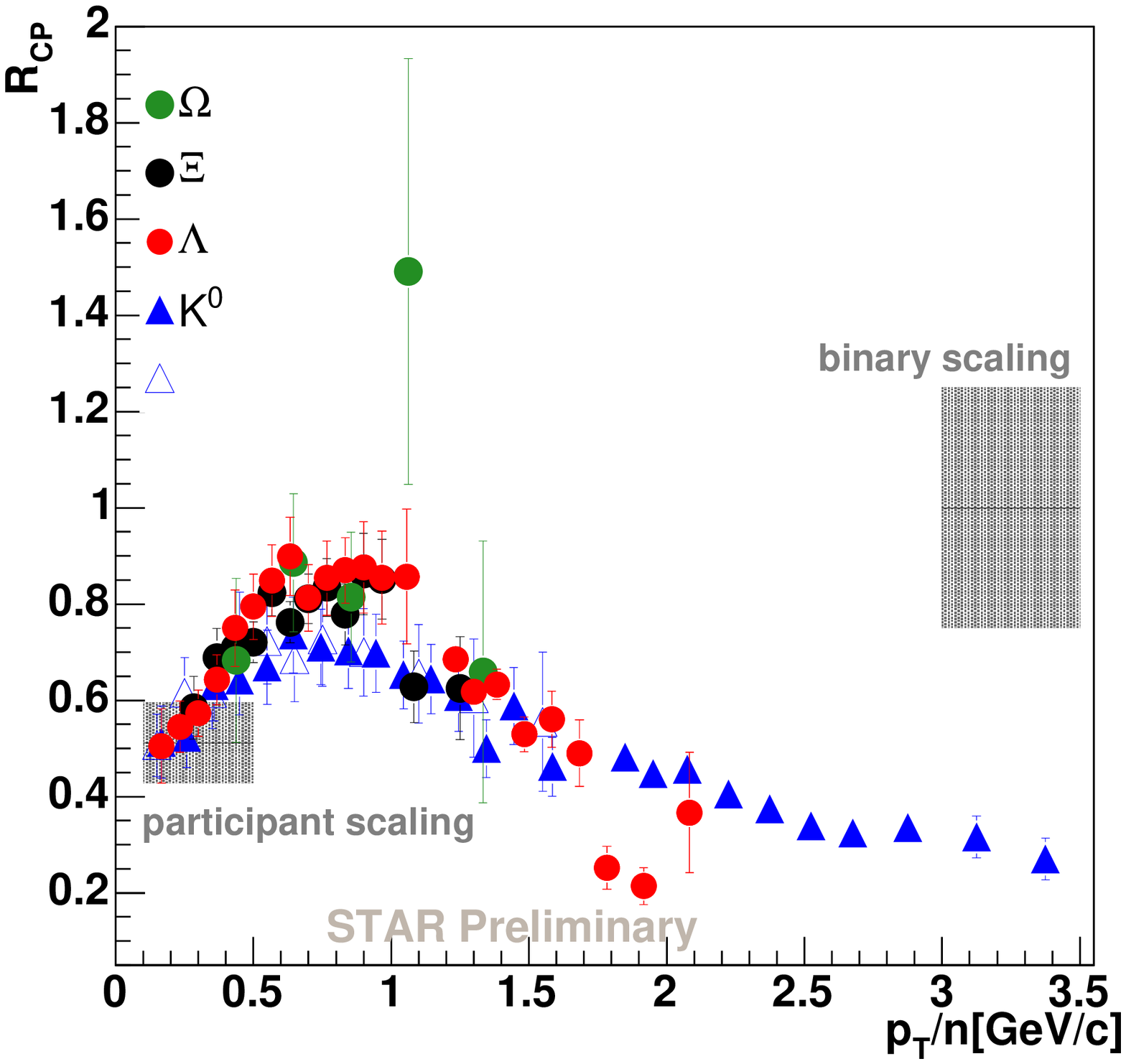} \\
 \mbox{\bf   (a) }&\mbox{\bf   (b)}
\end{array}$
\caption { {\bf   (a)}  $\rm R_{CP}$ vs $\rm p_{T}$ in \AuAu\
collisions at $\rm \sqrt{s_{NN}}=200 $ GeV. {\bf (b)}  $\rm
R_{CP}$ vs $\rm p_{T}/n$ in \AuAu\ collisions at $\rm
\sqrt{s_{NN}}=200$ GeV for $n=3$ for baryons and $n=2$ for mesons.
$\rm R_{CP}$ is calculated from 0-5\% and 40-60\% central \AuAu\
collisions. }
\label{fig:rcp}      
\end{figure}
 Nuclear modification factors for strange particles in \AuAu\
 collisions are presented in Figure~\ref{fig:rcp}-a. At high $\rm p_{T}$,
the ratios exhibit a suppression from binary scaling, attributed
to fast moving partons losing energy as they traverse a dense
medium. Hadron production  at high \pT\ does not follow binary
scaling and is roughly consistent with  participant scaling
\cite{Rcp}. The clear differences between baryons and mesons, is
believed to be due to hadron production through quark coalescence
at intermediate \pT \cite{matt}. For baryons and mesons, the
suppression sets in at a different \pT. Motivated by the
coalescence picture, Figure~\ref{fig:rcp}-b shows the $\rm R_{CP}$
ratio vs $ p_{T}/n$ for $  \sqrt {s_{NN}}=200 $ GeV, where n is
the number of valence quarks. Thus $ p_{T}/n$ represent the \pT\
of a valence quark. The baryon and meson difference sets in at the
same quark \pT, in agreement with the coalescence picture. This is
also observed for  $ \sqrt{s_{NN}}=62.4 $ GeV collisions
\cite{ref:salurQM04}.

Separation of the mesons and baryons is observed when the \pT\ of
the quark of the meson or the baryon is in the range of 0.8-1.2 GeV
which corresponds to a baryon $p_{T}$ range of 2.4-3.6 GeV and a
meson \pT\ range of 1.6-2.4 GeV. If the statistical fit parameters
from the ratios of the particle yields in these momentum ranges are
compared to the ones from the integrated momentum range, it might be
possible to determine the ranges of $T$ and \gs\ for which
coalescence and recombination are applicable. The yields are
extracted from the exponential distributions of the corrected \pT\
spectra in the given \pT\ ranges. The particle ratios from these sub
\pT\ ranges are given in Table~\ref{tabratio2}.

\begin{table}[hb]
\vspace*{-12pt} \caption[]{The particle ratios for $p_{T}$ of the
quark 0.8-1.2 GeV that corresponds to baryon $p_{T}$ of (2.4-3.6)
and meson $p_{T}$ of (1.6-2.4) in \AuAu\ collisions at
$\sqrt{s_{NN}}=200$ GeV. }\label{tabratio2} \vspace*{-14pt}
\begin{center}
\begin{tabular}{llllll}
\hline\\[-10pt]
Ratio & STAR data &~~~& Ratio & STAR data \\
\hline\\[-10pt]
$\pi^{-}/\pi^{+}$& $1.01 \pm 0.02$ & ~~~&$K^{-}/K^{+}$ & $0.96 \pm 0.03$\\
$\bar{p}/p$& $0.77 \pm 0.05$ &~~~ &$p/\pi^{-}$ & $0.137 \pm 0.013$\\
$\Lambda/\pi^{-}$& $0.156 \pm 0.015$ & ~~~&$\bar{\Lambda}/\Lambda$ & $0.72 \pm 0.024$\\
$\Xi^{-}/\pi^{-}$& $0.029 \pm 0.003$ & ~~~&$\bar{\Xi}^{+}/\Xi^{-}$ & $0.82 \pm 0.05$\\
$\bar{\Omega}/\Omega$& $1.01 \pm 0.08$ & & &\\
\hline
\end{tabular}
\end{center}
\end{table}

\begin{figure}[here!]
\centering $\begin{array}{cc} 
\hspace{-0.3in}
  \includegraphics[height=5.8cm,clip]{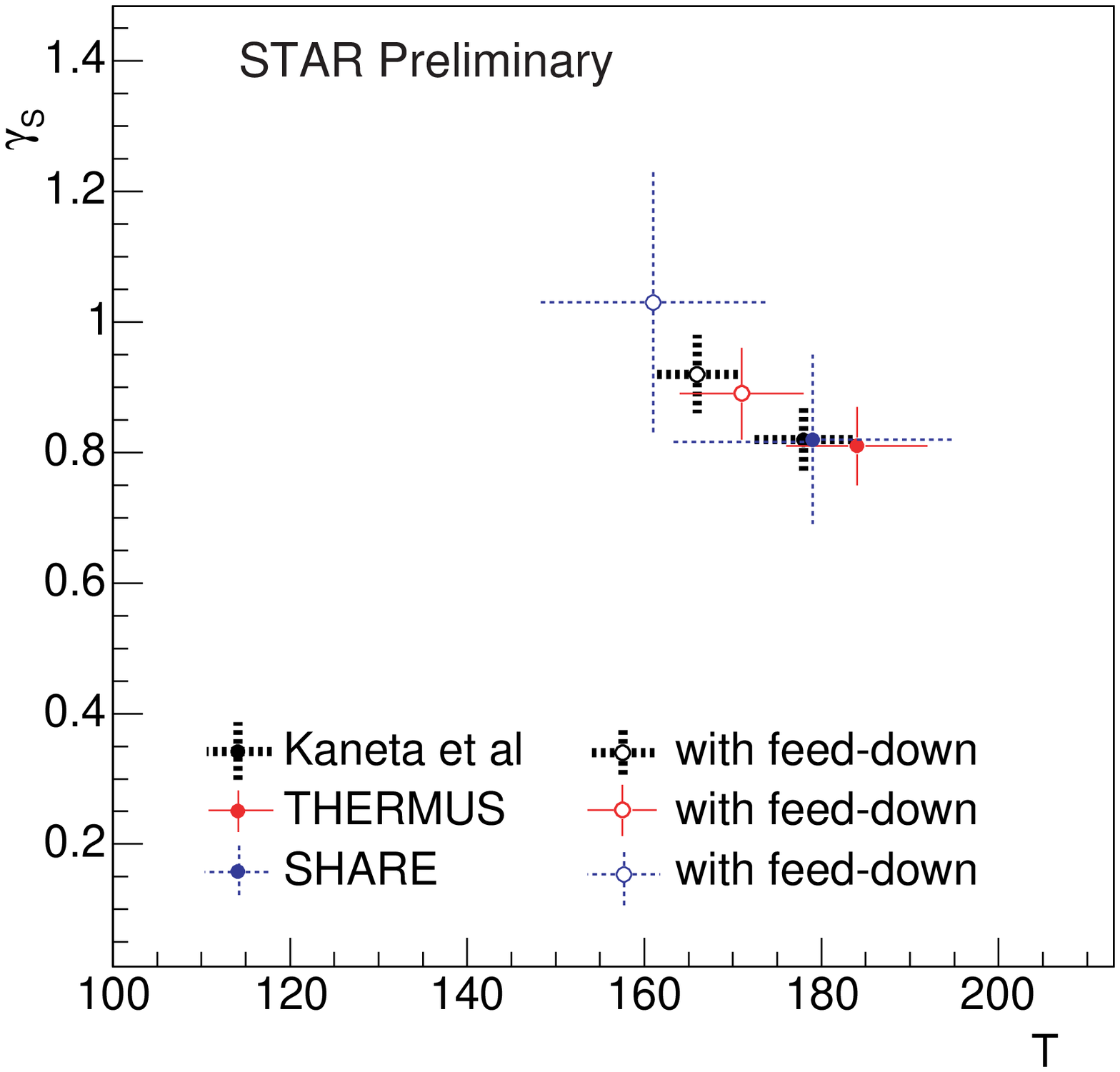} &
  \hspace{-0.1in}
\includegraphics[height=6cm,clip]{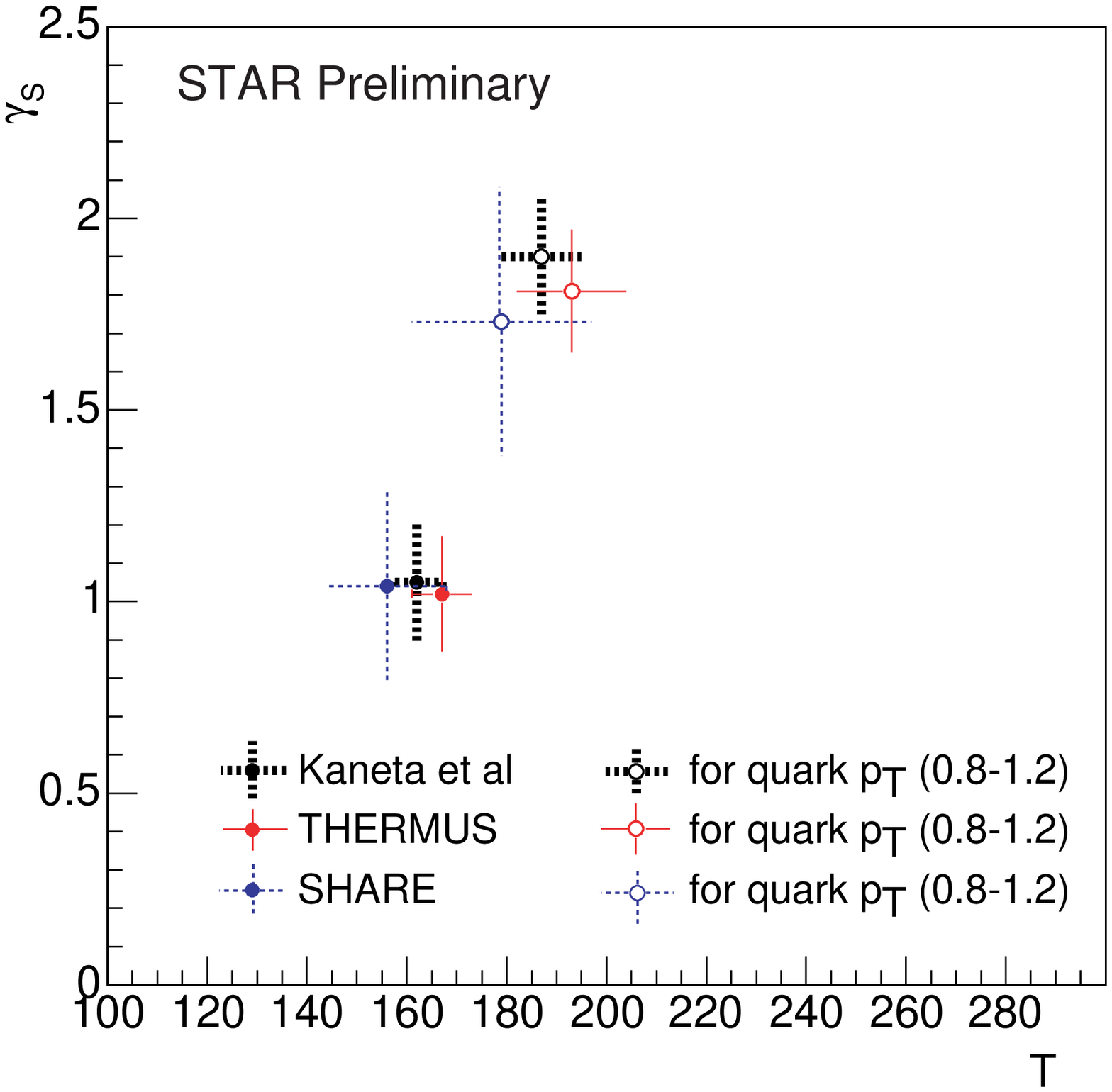} \\
 \mbox{\bf   (a) }&\mbox{\bf   (b)}
\end{array}$
\caption { {\bf   (a)} The \gs\ and the $T$ fit parameters of the
integrated ratios from the given statistical models in comparison to
the ones corrected for the weak decay feed-down. Input ratios of the
fits can be find in \cite{ref:olga}. {\bf (b)} The \gs\ and the $T$
fit parameters of the ratios from the quark \pT\ range of 0.8-1.2
GeV.}
\label{fig:thercomp}      
\end{figure}

A comparison of the $T$ and the \gs\ fit parameters are presented in
Figure~\ref{fig:thercomp}. The \pT\ integrated particle ratios that
are used in the fits for Figure~\ref{fig:thercomp}-a (closed
circles) can be found in Reference \cite{ref:olga}. A 40\% feed-down
correction is applied to the proton yields determined from the
$\Lambda$ yields for the weak feed-down corrected fits (open
circles). The effect of the weak feed-down correction can be seen as
an increase in the \gs\ and a reduction in the $T$. For any
conclusion made using the statistical model fits, it is essential to
remember that proper weak feed-down contributions plays a major role
in the final fit parameters.

The particle ratios presented in Table~\ref{tabratio2} are used in
the fits of Figure~\ref{fig:thercomp}-b (open circles) for
comparison with the \pT\ integrated ratios (solid circles). Since
there is no $K/\pi$ or $\Omega/\pi$ ratios, the fits are less well
constrained leading to larger errors as can be seen on the solid
circles in Figure~\ref{fig:thercomp}-b as compared to those on the
open circles in Figure~\ref{fig:thercomp}-a.  $T$ and $\gamma_{s}$
increase for the quark \pT\ range of 0.8-1.2 GeV relative to those
for the whole \pT\ range. Particles produced in this \pT\ range
possibly come from a hotter source.

\section{Conclusions}\label{concl}
The STAR experiment has collected tremendous amounts of data to
study strangeness production in heavy-ion collisions at a variety
of energies. The strange anti-particle to particle ratios show
that baryon transport is approximately independent of system size
at RHIC energies. Particle ratios in general can be used to
investigate the fireball properties with the help of statistical
models. We used several models to investigate the particle
production in \pp\ and \AuAu\ collisions at $\sqrt{s_{NN}}=200$
GeV. Comparisons of the various models and their corresponding
fits are also discussed. In these fits and model comparisons, the
importance of the weak decay feed-down corrections is highlighted.
$T$ is similar for \pp\ and \AuAu\ collisions at the same energy.
While $\gamma_{s}$  of \pp\ collisions at RHIC is smaller than the
one in \AuAu,  it is similar to that of $\rm Pb+Pb$ at SPS.

Baryon and meson suppression sets in at same quark \pT. Since
coalescence can partially explain the difference between baryons and
mesons in $\sqrt{s_{NN}}=200$ GeV collisions, the region where
coalescence dominates the particle production can be investigated
further with  statistical models. Where quark coalescence seems to
dominate (intermediate  \pT) $T$ and $\gamma_{s}$  show an increase.
This implies that particles in this range of \pT\ come from a hotter
source.

\bibliography{www27bib}
\bibliographystyle{lajolla2006}

\vfill\eject
\end{document}